\documentclass[twoside,11pt]{article}

\usepackage{tikz}
\usetikzlibrary{matrix}

\usepackage{bm}
\usepackage{jmlr2e}
\usepackage{amsmath}
\usepackage{algorithm, algorithmic}

\newcommand{\dataset}{{\cal D}}

\ShortHeadings{Financial Trading as a Game}{Huang, Chien-Yi}
\firstpageno{1}

\begin{document}

\title{Financial Trading as a Game: \\A Deep Reinforcement Learning Approach}

\author{\name Huang, Chien-Yi \email cyhuang.am03g@nctu.edu.tw \\
       \addr Department of Applied Mathematics\\
       National Chiao Tung University\\
       Hsinchu, Taiwan}
       
\editor{ABC XYZ}

\maketitle

\begin{abstract}
An automatic program that generates constant profit from the financial market is lucrative for every market practitioner. Recent advance in deep reinforcement learning provides a framework toward end-to-end training of such trading agent. In this paper, we propose an \textit{Markov Decision Process} (MDP) model suitable for the financial trading task and solve it with the state-of-the-art \textit{deep recurrent Q-network} (DRQN) algorithm. We propose several modifications to the existing learning algorithm to make it more suitable under the financial trading setting, namely  1. We employ a substantially small replay memory (only a few hundreds in size) compared to ones used in modern deep reinforcement learning algorithms (often millions in size.)  2. We develop an \textit{action augmentation} technique to mitigate the need for random exploration by providing extra feedback signals for \textit{all} actions to the agent. This enables us to use greedy policy over the course of learning and shows strong empirical performance compared to more commonly used $\epsilon$-greedy exploration. However, this technique is specific to financial trading under a few market assumptions.  3. We sample a longer sequence for recurrent neural network training. A side product of this mechanism is that we can now train the agent for every $T$ steps. This greatly reduces training time since the overall computation is down by a factor of $T$. We combine all of the above into a complete online learning algorithm and validate our approach on the spot foreign exchange market.
\end{abstract}


\section{Introduction}
In this paper we investigate the effectiveness of applying deep reinforcement learning algorithms to the financial trading domain. Financial trading, differs from the gameplay domain or robotics, posts some unique challenges. We point out some of them that we believe hold the key to successful application.

\subsection{The Financial Trading Task}
One way to describe the financial trading task is as the following:

\begin{center}
"An agent interacts with the market trying to achieve some intrinsic goal."
\end{center}

Note that the agent needs not to be human; algorithmic trading now accounts for large amount of trading activities in modern financial markets. Common interaction involves agent observing newly coming financial data or submitting new order to the exchange, etc. The intrinsic goal, say, for a hedge fund manager may be a risk-adjusted measure e.g. trying to reach a 15\% annual return target under a specified volatility threshold. The goal for a naive trader may be simply to pursuit highest profit without properly take into account the risk incurred. An extreme example is that an individual that trades only for the "gambling sensation" and doesn't care about the financial market at all.

Although the above description is quite general, there are some characteristics of such task:
\begin{enumerate}
\item The agent interacts with the financial market at \textit{discrete} time steps even though the time steps may be extremely close, say, in the high frequency trading, trading decisions can be made in the matter of milliseconds.
\item There is a set of legal \textit{actions} an agent can apply to the market from naively submitting market orders with a fixed position size to submitting a fully specified limit order.
\item The financial market produces new information available to the agent at each time step enables the agent to make trading decisions. However, the agent doesn't have full clue on how the data is generated.
\item The agent has the potential to \textit{alter}, although without full control over, the financial market if it is powerful enough. Hence it is not entirely realistic to consider the market to be fully exogenous to the agent.
\end{enumerate}

With these characteristics, we desire a unified framework for training such agent. This is part of the motivation behind this thesis.

\subsection{Motivation}
With many successful stories of deep reinforcement learning, a natural question to ask is:

\begin{center}
"Can an artificial agent learn to trade successfully?"
\end{center}

\textit{Success} is defined in terms of the degree the agent is reaching its intrinsic goal. One of the most fundamental hypotheses of reinforcement learning is that \textit{goals} of an agent can be expressed through maximizing long-term future rewards. \textit{Reward} is a single scalar feedback signal that reflects the "goodness" of an agent's action in some state. This is called the \textit{reward hypothesis}.

\begin{definition} {(Reward Hypothesis)}
\textit{All} goals can be described by maximization of expected
future reward.
\end{definition}

The four characteristics mentioned above resembles that of \textit{reinforcement learning}. A branch of machine learning that studies the science of sequential decision-making. Reinforcement learning has recently received a considerable amount of attention due to solving challenging control tasks that are infeasible before. The motivation behind this thesis is therefore to see if the recently proposed techniques migrate to the financial trading task and to see how far we can go with these techniques.

\subsection{Challenges}
We identify four major challenges to applying reinforcement learning to financial trading:

\begin{enumerate}
\item \textbf{Lack of baseline.} With a large body of work published on applying deep reinforcement learning to video gameplay and robotics. There is relatively little work on how to apply the same algorithm to financial trading \cite{li2017deep}. There is no clear baseline nor a suitable MDP model, network architecture or a set of hyperparameters can be followed on early stage experiments.

\item \textbf{Data quality and availability.} Financial data are difficult to obtain in high resolution. Usually only open, high, low and close prices (OHLC) data are freely accessible which, may not be sufficient to produce successful trading strategy. The financial time series itself is non-stationary, posting challenges to standard, gradient-based learning algorithms.

\item \textbf{Partially observability of financial markets.} No matter how "complete" our input state is, there will always be a degree of unobservability in the financial market. We are unable to observe, for every market participator, their consensus on current market condition.

\item \textbf{Exploration and exploitation dilemma.} Despite the sophistication of modern deep reinforcement learning algorithms, usually a naive \textit{exploration policy} is used. For example, $\epsilon$-greedy exploration in valued-based methods and Boltzmann exploration in policy-based methods \cite{sutton1998reinforcement}. This is infeasible in financial trading setting since random exploration would inevitably generate huge amount of transaction costs and hurt performance.
\end{enumerate}

\subsection{Contributions}
The contributions of this thesis are three-fold:

\begin{enumerate}
\item We propose a Markov decision process (MDP) model for general signal-based financial trading task solvable by state-of-the-art deep reinforcement learning algorithm with publicly accessible data only. The MDP model is easily extendable with more sophisticated input features and more complex action spaces with minimal modifications to model architecture and learning algorithm.

\item We modify the existing deep recurrent Q-network algorithm in a way that it's more suitable for the financial trading task. This involves using a substantially smaller replay memory and sampling a longer sequence for training. We are surprised by the above two discoveries since in deep reinforcement learning, usually a large replay memory is used and length of the sampled sequences are usually only a few time steps long. We also discover workable hyperparameters for the DRQN algorithm able to solve the financial trading MDP through random search. We also develop a novel \textit{action augmentation} technique to mitigate the need for random exploration in the financial trading environment.

\item We achieve positive return on 12 different currency pairs including major and cross pairs under transaction costs. To the author's best knowledge, this is the first successful application on real financial data using pure deep reinforcement learning techniques. Numerical results presented in this paper can serve as benchmarks for future studies.
\end{enumerate}

This thesis is structured as follows: in section 2, we give a detailed description on the proposed method including data preparation, feature extraction, model architecture and learning algorithm. In section 3, we combine all of the proposed techniques into a single online learning algorithm. In section 4, we evaluate our algorithm on the spot foreign exchange market and provide numerical results.


\section{Method}
In this section we provide detailed description on the proposed MDP model, model architecture as well as the learning algorithm.

\subsection{Data Preparation and Feature Extraction}
We download tick-by-tick forex data from \texttt{TrueFX.com} from January 2012 to December 2017. We pick 12 currency pairs, namely AUDJPY, AUDNZD, AUDUSD, CADJPY, CHFJPY, EURGBP, EURJPY, EURUSD, GBPJPY, GBPUSD, NZDUSD and USDCAD. For diversity, both major and cross pairs are included. We then resample the data into 15-minute intervals with open, high, low, close prices and tick volume. The reason for choosing forex over other asset classes is the ease of accessing high resolution data, often at very low or no cost.

\subsection{Financial Trading MDP}
In this section we give definitions of the state space, action space and reward function of the financial trading MDP.

\subsubsection{State Space $\in\mathbb R^{198}$}
The state representation is a 198-dimensional vector consists of the following three parts:
\begin{itemize}
\item \textbf{Time feature} $\in \mathbb R^3$ \\
Since the foreign exchange market has the longest opening hours among all financial markets. In order for our agent to differentiate different market sessions, we add the \textit{minute}, \textit{hour} and \textit{day of week} of the current time stamp to be part of the state representation. This is encoded via a sinusoidal function
\[ \sin\left(2\pi\,\frac{t}{T}\right) \]
where $t$ is the current value (zero-based numbering) and $T$ is number of possible values for $t$.

\item \textbf{Market feature} $\in \mathbb R^{16\times 12}$ \\
We extract 16 features from OHLCV data containing 8 most recent log returns on both closing price and tick volume. A running Z-score normalization of period 96 is then applied to each dimension of the 16 input features. We also clip the value by 10 after normalization to eliminate outliers. We utilize price features from all 12 currency pairs in the hope that deep neural networks can extract useful intermarket features from the data.
\item \textbf{Position feature} $\in \mathbb R^3$ \\
The agent's current position is encoded via a 3-dimensional one-hot vector indicates whether the current position is of -1, 0 or +1 unit, e.g. if the current position is of +1 unit, the encoding would be $[0, 0, 1]$.
\end{itemize}

\subsubsection{Action Space}
We adopt a simple action set of three values \{-1, 0, 1\}. Position reversal is allowed (results in double amount of transaction costs). Note that when the current position is +1 and the agent again outputs +1 at the next time step, no trading action will be executed. This sometimes refers to as \textit{target orders} where the output indicates the target position size, not the trading decision itself. This simplifies the action space definition and makes the implementation easier.

\subsubsection{Reward Function}
We define the reward function as portfolio log returns at each time step, i.e.
\begin{align}
r_t = \log\left(\frac{v_t}{v_{t-1}}\right)
\label{eq:reward}
\end{align}
where $v_t$ is the portfolio value (account balance plus unrealized PnL from open positions). With the above definition, the portfolio value $v_t$ satisfies a simple recursive relation
\begin{align}
 v_t = v_{t-1} + a_t \cdot c \cdot (c_t - o_t) - d_t 
\end{align}
where $a_t$ is the output action, $c$ is the (constant) trade size, $o_t$, $c_t$ are the current open, close prices and $d_t$ is the commission term. The commission $d_t$ is computed by
\begin{align}
d_t = c \cdot |a_t - a_{t-1}| \cdot \text{spread}.
\end{align}

We use spread as a principled way of measuring the cost for making trading decisions. The spread we consider here, unlike real spreads, is kept fixed along the course of learning. This is for the ease of comparing different currency pairs since the width of the spread varies from pair to pair.

Defining the reward function this way, the return $G_t$ has a nice interpretation as the future discounted log returns. When facing action selection, the agent is effectively picking actions with the highest log returns. We prefer log returns over arithmetic returns as they are additive which is more natural in the RL setting.

\subsection{Fully Exploit with Action Augmentation}
Random exploration is unsatisfying in the financial trading setting since transaction costs occur with a change of position. We propose a simple technique to mitigate the need for exploration by providing the agent with reward signal for \textit{every} action. This is possible since the reward is easily computable after the price at the current timestep is observed using equation (\ref{eq:reward}). For example, if the unrealized PnL for the current step is +10 after we execute action +1, then we immediate know that if we were to execute action -1, we would get a reward of -10 and 0 for action 0. Therefore the portfolio value $v_t$ can be computed (therefore the reward signal) for \textit{all} actions.

On the other hand, the only part of the state that would be altered if we were to take other actions is the agent's \textit{position}. This is known as the \textit{zero market impact hypothesis} which states that the action taken from the market participator has no influence on the current market condition. We also assume order issued by the agent always executes at the next opening price. That is, we always know the position for the next step if the output action is determined.

Now we are able to update Q-values for \textit{all} actions. We write down a novel loss function in vector form called \textit{action augmentation loss},
\begin{align}
\mathcal L(\theta) &= \mathbb E_{(s, \bm a, \bm r, \bm{s'})\sim\mathcal D} \left[\Vert\textbf r + \gamma Q_{\theta^-}(\bm{s'}, \arg\max_{a'} Q_\theta(\bm s', a')) - Q_\theta(s, \bm a)\Vert^2\right] \label{AA_Loss} \\
\theta &\leftarrow \theta - \alpha \nabla_\theta \mathcal L(\theta) \label{AA_Update}
\end{align}
where $Q_{\theta^-}$ denotes the target network.

\subsection{Model Architecture}
We use a four-layered neural network as function approximator to represent the optimal action-value function $q_*$. The first two are linear layers with 256 hidden units and ELU \cite{clevert2015fast} activation. The third layer is an LSTM layer with the same size. The fourth layer is another linear layer with 3 output units. The network is relatively small with approximately 65,000 parameters.

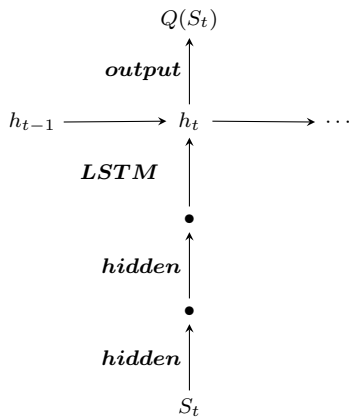
\begin{figure}[h]
\begin{center}
\begin{tikzpicture}
\scriptsize
	 \matrix (m) [matrix of math nodes,row sep=3em,column sep=4em,minimum width=2em]
  		{
        		& Q(S_t) & \\
        		h_{t-1} & h_t & \cdots \\
        		& \bullet & \\
        		& \bullet & \\
	    		& S_t & \\};
	    \path[-stealth]
	    (m-2-1) edge node [above] {} (m-2-2)
	    (m-2-2) edge (m-2-3)
	    (m-2-2) edge node [left] {$\bm{output}$} (m-1-2)
	    (m-3-2) edge node [left] {$\bm{LSTM~~~}$} (m-2-2)
	    (m-4-2) edge node [left] {$\bm{hidden}$} (m-3-2)
	    (m-5-2) edge node [left] {$\bm{hidden}$} (m-4-2);
		\end{tikzpicture}
\end{center}
\caption{Model architecture.}
\end{figure}

\subsubsection{Weight Initialization}
Weight initialization is crucial for successful training of deep neural networks. We follow initialization scheme presented in \cite{he2015delving} for weight matrices in both hidden layers and input-to-hidden layer in LSTM. We follow \cite{le2015simple} to initialize all hidden-to-hidden weight matrices to be identity. We set all biases in the network to be zero except for the forget gate in the LSTM which are set to be 1. We sparsely initialize the output layer weight matrix with Gaussian distribution $\mathcal N(0, 0.001)$.


\subsection{Training Scheme}
In this section we combine all of the above and present a complete learning algorithm that we will evaluate in section 5 on the spot foreign exchange market.

\subsubsection{Modified Training Scheme}
After some experiment with the original updating scheme for DRQN, we proposed the following modifications:
\begin{enumerate}
\item We discover that using a relatively small replay memory is more effective. It is different from the "common knowledge" in value-based deep reinforcement learning where large replay memories (often millions in size) are used. This makes intuitive sense since in financial trading recent data points are more important than those from the far past. The performance decreases if we enlarge the replay memory.
\item We sample a longer sequence from the replay memory than the number of steps used in the DRQN paper. The reason behind this modification is that, a successful trading strategy involves opening a position at the right time and \textit{holding} the position for a sufficiently long period of time then exiting the position. Sampling a short sequence can't effectively train the network to learn the desired long-term dependency.
\item We find that it is unnecessary to train the network for each step since we are sampling a longer sequence. Hence we only train the network for every $T$ time steps. This significantly reduces computation since the number of backward passes are reduced by a factor of $T$. This is also beneficial for real-time trading since trading decision can be carried out with low latency and training can be deferred after market close.
\end{enumerate}

\subsubsection{A Complete Online Learning Algorithm}
We adopt the above updating scheme to the original DRQN algorithm and propose a complete online learning algorithm that we will evaluate in the next section. We discard the common forwalk-walk optimization process that involves slicing the dataset into consecutive training and testing sets. Since each training set constructed this way are largely overlapped, strong overfitting is observed in our early stage experiments. We therefore optimize our network in a purely online fashion that most resembles real-time trading. We term the resulting algorithm \textit{financial deep recurrent Q-network} (Financial DRQN).

\begin{algorithm}[H]
\begin{algorithmic}[1]
	\STATE Initialize $T\in\mathbb N$, recurrent Q-network $Q_\theta$, target network $Q_{\theta^-}$ with $\theta^- = \theta$, \\dataset $\mathcal D$ and environment $E$, $steps = 1$
    \STATE Simulate env $E$ from dataset $\mathcal D$
	\STATE Observe initial state $s$ from env $E$
    \FOR {each step}
    \STATE $steps \leftarrow steps + 1$
    \STATE Select greedy action w.r.t. $Q_\theta(s, a)$ and apply to env $E$
    \STATE Receive reward $r$ and next state $s'$ from env $E$
    \STATE Augment actions to form $\mathcal T = (s, \bm a, \bm r, \bm{s'})$ and store $\mathcal T$ to memory $\mathcal D$
    \IF {$\mathcal D$ is filled and $steps \mod T = 0$}
    \STATE Sample a sequence of length $T$ from $\mathcal D$
    \STATE Train network $Q_\theta$ with equation (\ref{AA_Loss}) and (\ref{AA_Update})
    \ENDIF
    \STATE Soft update target network $\theta^- \leftarrow (1-\tau)\theta^- + \tau \theta$
	\ENDFOR
\end{algorithmic}
\caption{Financial DRQN Algorithm}
\label{FDRQN}
\end{algorithm}

In practice, we find it useful to implement a simple OpenAI Gym-like environment \cite{brockman2016openai} for training. Since most open source backtesting engine is difficult to work with under the RL paradigm.

\section{Experiment}
In this chapter we present numerical results for the financial DRQN algorithm on 12 currency pairs, test the algorithm against different spread settings and investigate the usefulness of the proposed action augmentation technique.

\subsection{Hyperparameters}
In this section, we validate our approach on the spot foreign exchange market. We believe our method can be extended to other financial markets with minimal modification. Hyperparameters used in Algorithm \ref{FDRQN} are listed below which are quite standard in modern deep reinforcement learning literature. Hyperparameters and model architecture are kept fixed across all experiments.

\begin{center}
  \begin{tabular}{r | r}
    Hyperparameters & Value \\
    \hline
    Learning timestep $T$ & 96 \\
    Replay memory size $N$ & 480 \\
    Learning rate & 0.00025 \\
    Optimizer & Adam\footnotemark[1] \\
    Discount factor & 0.99 \\
    Target network $\tau$ & 0.001 \\
  \end{tabular}
\end{center}
\footnotetext[1]{The Adam optimizer \cite{kingma2014adam}}

We did not do an exhaustive search over hyperparameters but stick to ones that shows good empirical results. 

\subsection{Simulation Result}
We need additional parameters for trading simulation. The parameters are mainly used to compute trading statistics such as annual return and Sharpe ratio. Parameters used are listed below and kept fixed for every simulation.

\begin{center}
  \begin{tabular}{r | r}
    Simulation Parameters & Value \\
    \hline
	Initial cash & 100,000\footnotemark[2] \\
	Trade size & 100,000 \\
    Spread (bp\footnotemark[3]) & 0.08 \\
    Trading days & 252 days/year \\    
  \end{tabular}
\end{center}
\footnotetext[2] {Initial cash is 100,000 in base currency.}
\footnotetext[3] {We keep bp to be 0.0001 for non-JPY quoted currencies and 0.01 for JPY-quoted currencies.}

Below we present numerical results for 12 currency pairs. We consider two baselines: buy-and-hold and "sell-and-hold" since some of the currency pairs show constant down trend throughout the test period. The one producing larger gain is used as baseline. Profit and loss are reported in terms of cumulative percentage returns. Every experiment is carried out for 5 times. This serves as a "robustness test" for the proposed approach. Equity curve averaged over 5 runs are plotted in blue curve with one standard deviation range in shaded area.

Table \ref{table:sim_results} summarizes the performance for each currency pair. Annual return and risk-adjusted metrics are calculated by first calculating the daily return\footnotemark[4] and then annualized by multiplying by factor 252($\sqrt{252}$ for Sharpe and Sortino ratios.) Baselines are also provided in the parenthesis for annual return. Maximum drawdown (MDD) and log return correlation between the baseline is also computed using daily return.
\footnotetext[4] {We group one-step PnL into consecutive 96 steps to form the "daily" PnL.}

\begin{table}[h]
\begin{center}
\begin{tabular}{r r r r r r r}
\hline
& Net Profit & Return & Sharpe & Sortino & MDD & Corr \\
\hline
GBPUSD & 93876.40 & 16.2\% (-3.5\%) & 1.5 & 2.5 & -8.63\% & -0.09 \\
EURUSD & 55046.00 & 9.5\% (-1.6\%) & 1.0 & 1.6 & -11.76\% & 0.01 \\
AUDUSD & 85658.40 & 14.8\% (-4.2\%) & 1.7 & 2.7 & -6.96\% & 0.02 \\
NZDUSD & 98984.80 & 17.1\% (-1.2\%) & 2.2 & 4.0 & -4.17\% & -0.04 \\
USDCAD & 71585.40 & 12.2\% (4.0\%) & 1.4 & 2.5 & -6.21\% & 0.11 \\
EURGBP & 61927.80 & 12.8\% (1.1\%) & 1.8 & 3.5 & -5.51\% & -0.21 \\
AUDNZD & 260776.20 & 34.3\% (-2.8\%) & 5.7 & 12.4 & -1.21\% & 0.02 \\
CADJPY & 8923129.40 & 20.4\% (3.2\%) & 1.8 & 3.1 & -25.24\% & 0.20 \\
AUDJPY & 11404412.00 & 25.1\% (2.0\%) & 2.0 & 3.3 & -11.69\% & 0.18 \\
CHFJPY & 28816485.20 & 60.8\% (7.0\%) & 3.1 & 6.3 & -7.71\% & 0.31 \\
EURJPY & 13576373.50 & 23.6\% (6.1\%) & 1.9 & 3.2 & -12.90\% & 0.18 \\
GBPJPY & 26931635.80 & 39.0\% (4.7\%) & 2.9 & 5.8 & -7.73\% & -0.07 \\
\hline
\end{tabular}
\caption{Annualized simulation results}
\label{table:sim_results}
\end{center}
\end{table}

Additional statistics on the overall trading activities is summarized in Table \ref{table:trade_stats}. We discover that the agent favors higher win rates (around 60\%) while maintaining a roughly equal average profit and loss per trade (about 2 basic points in difference). The trading expectation is calculated using the win rate and average PnL. Trading frequency is calculated by dividing the length of the data by total number of trades.

\begin{table}[h]
\begin{center}
\begin{tabular}{r r r r r r r}
\hline
& Num Trades & Win Rate & Avg Profit & Avg Loss & Expect & Freq \\
\hline
GBPUSD & 33133 & 57.2\% & 70.25 & -87.33 & 2.83 & 4.22 \\
EURUSD & 31215 & 57.2\% & 60.67 & -77.12 & 1.76 & 4.48 \\
AUDUSD & 31263 & 57.2\% & 54.52 & -66.6 & 2.74 & 4.47 \\
NZDUSD & 32382 & 59.6\% & 52.17 & -69.34 & 3.06 & 4.32 \\
USDCAD & 26636 & 57.7\% & 63.46 & -80.16 & 2.71 & 5.25 \\
EURGBP & 32032 & 61.2\% & 37.76 & -54.58 & 1.93 & 4.36 \\
AUDNZD & 38173 & 63.2\% & 49.93 & -67.18 & 6.83 & 3.66 \\
CADJPY & 26332 & 59.6\% & 6410.1 & -8612.01 & 340.1 & 5.31 \\
AUDJPY & 26638 & 60.7\% & 7092.02 & -9883.08 & 428.92 & 5.25 \\
CHFJPY & 32089 & 61.5\% & 7287.77 & -9294.91 & 898.92 & 4.36 \\
EURJPY & 30509 & 61.5\% & 7483.41 & -10801.23 & 445.0 & 4.58 \\
GBPJPY & 31204 & 60.8\% & 10791.52 & -14503.05 & 864.67 & 4.48 \\
\hline
\end{tabular}
\caption{Trading statistics}
\label{table:trade_stats}
\end{center}
\end{table}

\subsection{Effect of the Spread}

\begin{table}[h]
\begin{center}
\begin{tabular}{r r r r r}
\hline
& 0.08 bp & 0.1 bp & 0.15 bp & 0.2 bp \\
\hline
GBPUSD & 16.2\% & \textbf{18.8\%} & 6.1\% & 6.7\%\\
EURUSD & \textbf{9.5\%} & 5.8\% & 0.1\% & 1.1\%\\
AUDUSD & \textbf{14.8\%} & 10.0\% & 7.3\% & 5.2\%\\
NZDUSD & \textbf{17.1\%} & 14.2\% & 12.4\% & 4.2\%\\
USDCAD & \textbf{12.2\%} & 9.0\% & 6.9\% & -3.4\%\\
EURGBP & \textbf{12.8\%} & 3.8\% & -0.2\% & -3.8\%\\
AUDNZD & 34.3\% & \textbf{35.9\%} & 29.9\% & 23.4\%\\
CADJPY & 20.4\% & \textbf{32.4\%} & 18.9\% & 14.8\%\\
AUDJPY & 25.1\% & \textbf{26.4\%} & 15.3\% & 10.2\%\\
CHFJPY & 60.8\% & \textbf{79.8\%} & 56.1\% & 43.5\%\\
EURJPY & 23.6\% & \textbf{35.6\%} & 17.2\% & 15.2\%\\
GBPJPY & 39.0\% & \textbf{44.4\%} & 31.0\% & 27.0\%\\
\hline
 & 23.8\% & \textbf{26.3\%} & 16.7\% & 11.9\% \\
\hline
\end{tabular}
\caption{Annualized return under different spreads.}
\label{table:effect_spread}
\end{center}
\end{table}

Since spread is the only source for market friction, it is meaningful to examine algorithm performance under various spread settings. We experiment with spread levels 0.08, 0.1, 0.15 and 0.2 basic points\footnotemark[5] and discover the following facts:
\footnotetext[5] {Spread levels are taken from a leading online breaker \textit{Interactive Brokers}.}
\begin{enumerate}
\item Generally, wider spread results in worse performance. It fits our intuition since the transaction costs paid are proportional to the width of the spread.
\item The agent stays profitable for most of the currency pairs under 0.15 basic points of spread. Profitable strategies cannot be discovered under 0.2 basic point of spread for currency pair USDCAD and EURGBP.
\item An interesting discovery is that \textit{wider spread does not always lead to worse performance.} For some of the JPY-quoted currency pairs the performance actually enhanced. We deem that a slightly wider spread forces the agent to locate a more reliable strategy that is more robust under market change.
\end{enumerate}

\subsection{Effectiveness of Action Augmentation}
We investigate how useful is the action augmentation technique by comparing it to a traditional $\epsilon$-greedy policy with $\epsilon=0.1$. With action augmentation, performance improves and standard deviation narrows, showing the algorithm is more robust and reliable. Table \ref{table:action_augment} lists performances for both $\epsilon$-greedy policy and action augmentation. We gain an additional 6.4\% annual return in average when we use action augmentation. 

\begin{table}[h]
\begin{center}
\begin{tabular}{r r r r}
\hline
& $\epsilon$-greed & Act Aug & Gain \\
\hline
GBPUSD & 13.7\% & 16.2\% & 2.5\% \\
EURUSD & 7.1\% & 9.5\% & 2.4\% \\
AUDUSD & 6.4\% & 14.8\% & 8.4\% \\
NZDUSD & 9.5\% & 17.1\% & 7.6\% \\
USDCAD & -4.1\% & 12.2\% & 16.3\% \\
EURGBP & 7.1\% & 12.8\% & 5.8\% \\
AUDNZD & 28.1\% & 34.4\% & 6.3\% \\
CADJPY & 17.9\% & 20.4\% & 2.5\% \\
AUDJPY & 20.3\% & 25.0\% & 4.8\% \\
CHFJPY & 57.0\% & 60.8\% & 3.8\% \\
EURJPY & 15.0\% & 23.6\% & 8.6\% \\
GBPJPY & 30.9\% & 39.0\% & 8.1\% \\
\hline
 & 17.4\% & 23.8\% & \textbf{6.4\%} \\
\hline
\end{tabular}
\caption{Annualized returns with and without action augmentation.}
\label{table:action_augment}
\end{center}
\end{table}

\section{Conclusion}
We conclude and point future directions for this thesis in this chapter.

\subsection{Achievements}
The achievements of this thesis can be summarized as follows:
\begin{enumerate}
\item We propose an MDP model for signal-based trading strategies that is flexible to future extensions with minimal modifications to model architecture and learning algorithm.

\item We modify the existing deep recurrent Q-network learning algorithm to make it more suitable in the financial trading setting. Especially, we propose an action augmentation technique to mitigate the need for random exploration. We also use a substantially smaller replay memory compared to ones used in value-based deep reinforcement learning.

\item We give empirical results on the proposed algorithm for 12 currency pairs and achieve positive results under most simulation settings. To the author's best knowledge, this is the first positive result achieved by pure deep reinforcement learning algorithm under transaction costs. Strategies discovered by the agent exhibit low or no correlation between baselines.

\item We discover a counter-intuitive fact that a slightly increased spread leads to better overall performance. This phenomena is observed for over half of the currency pairs. We think a slightly higher spread forces the agent to discover more robust and reliable trading strategies over the learning process. However, further widening the spread destroys performance.
\end{enumerate}

\subsection{Future Work}
There are many potentials for future improvements on the proposed method. We list some of them that we believe are most important and interesting:

\begin{enumerate}
\item Expand state space and action space. We may augment more input features such as price data from other markets (even ones seemly unrelated at first glance), macro data (released news from politics and economics, fundamental data such as economic indices). For the action space, we may give the agent more freedom when making trading decisions such as deciding how much to invest (i.e. the position size) or even posting limit orders. This would requires a more complex action space and a careful output \textit{representation} of an action.

\item Apply reinforcement learning to different trading scenarios, e.g. high frequency trading, pair trading or long-term equity investment. This is a further test for robustness of our method. A portfolio combining many different strategies can be created to suit investors' needs.

\item Make use of \textit{distributional reinforcement learning} \cite{bellemare2017distributional} to take risk-adjusted actions. In distributional reinforcement learning, rather than learning the \textit{expected} return $\mathbb E[Q(s, a)]$, the entire \textit{distribution} over $Q(s, a)$ is learned. This is possible due to a distributional variant of the Bellman equation,
\[
Q(s, a) \overset{D}{=} R(s, a) + \gamma Q(S', A').
\]
In this thesis, we choose actions solely to maximize the expected return. That is, we blindly maximize profit without taking risk into account. This is unsatisfying since it is clear that we would prefer an trading decision that comes with lower variance although it may be less profitable. Since the entire distribution is learned, we are able to choose action with the highest expected Q-value \textit{and} the lowest standard deviation of the Q-value, i.e.
\[
a = \arg\max_{a\in\mathcal A} \frac{\mathbb E[Q]}{\sqrt{\text{Var}[Q]}}.
\]
This way, we choose actions with the highest \textit{Sharpe ratio} and the strategy would be more suitable for modern investors.
\end{enumerate}

\newpage
\section*{Appendix}
In this appendix we provide equity curves for all experiments done in section 4. ~\\

\begin{figure}[h]
\begin{center}
\includegraphics[width=.9\linewidth]{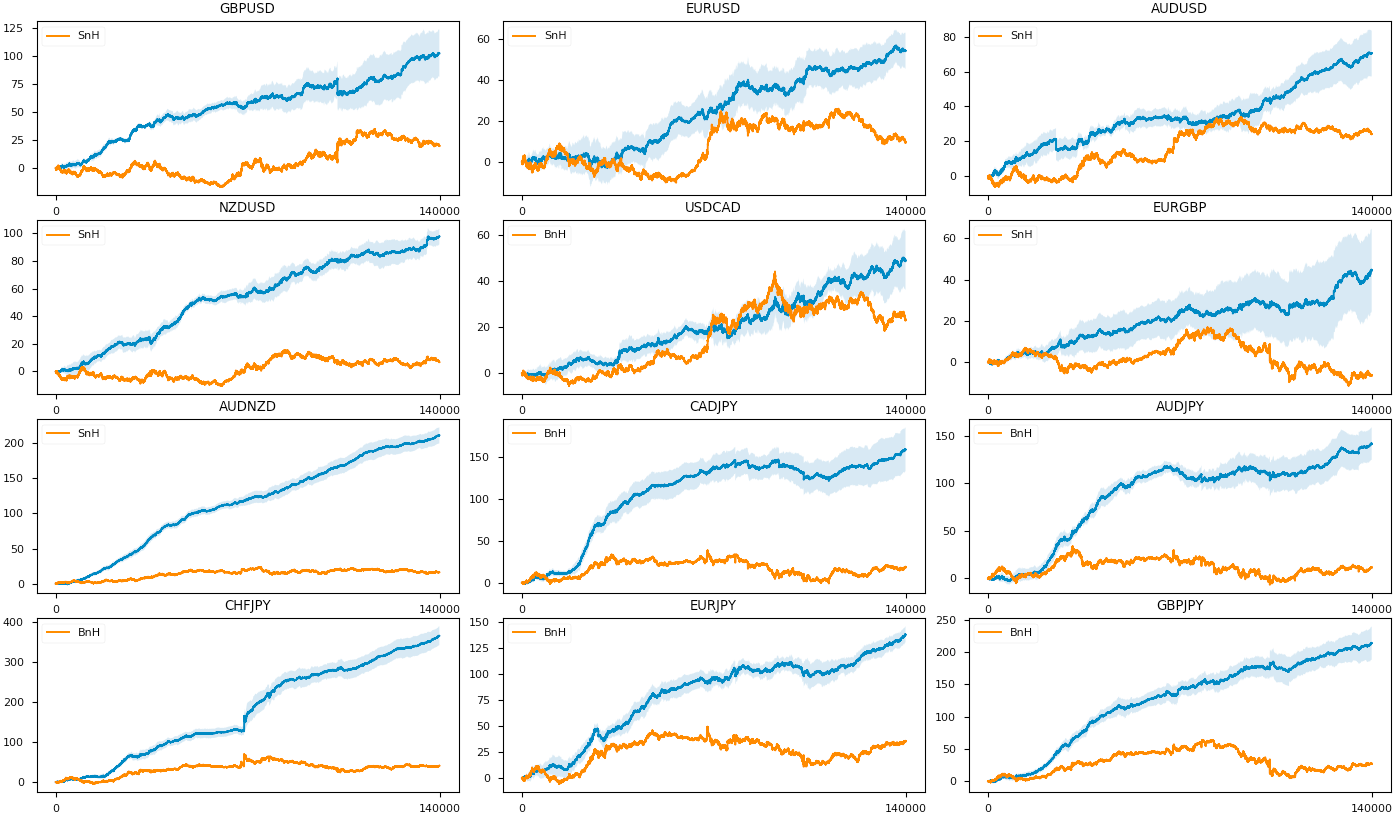}
\end{center}
\caption{Performance under 0.08 basic points of spread.}
\end{figure}

\begin{figure}[h]
\centering\includegraphics[width=.9\linewidth]{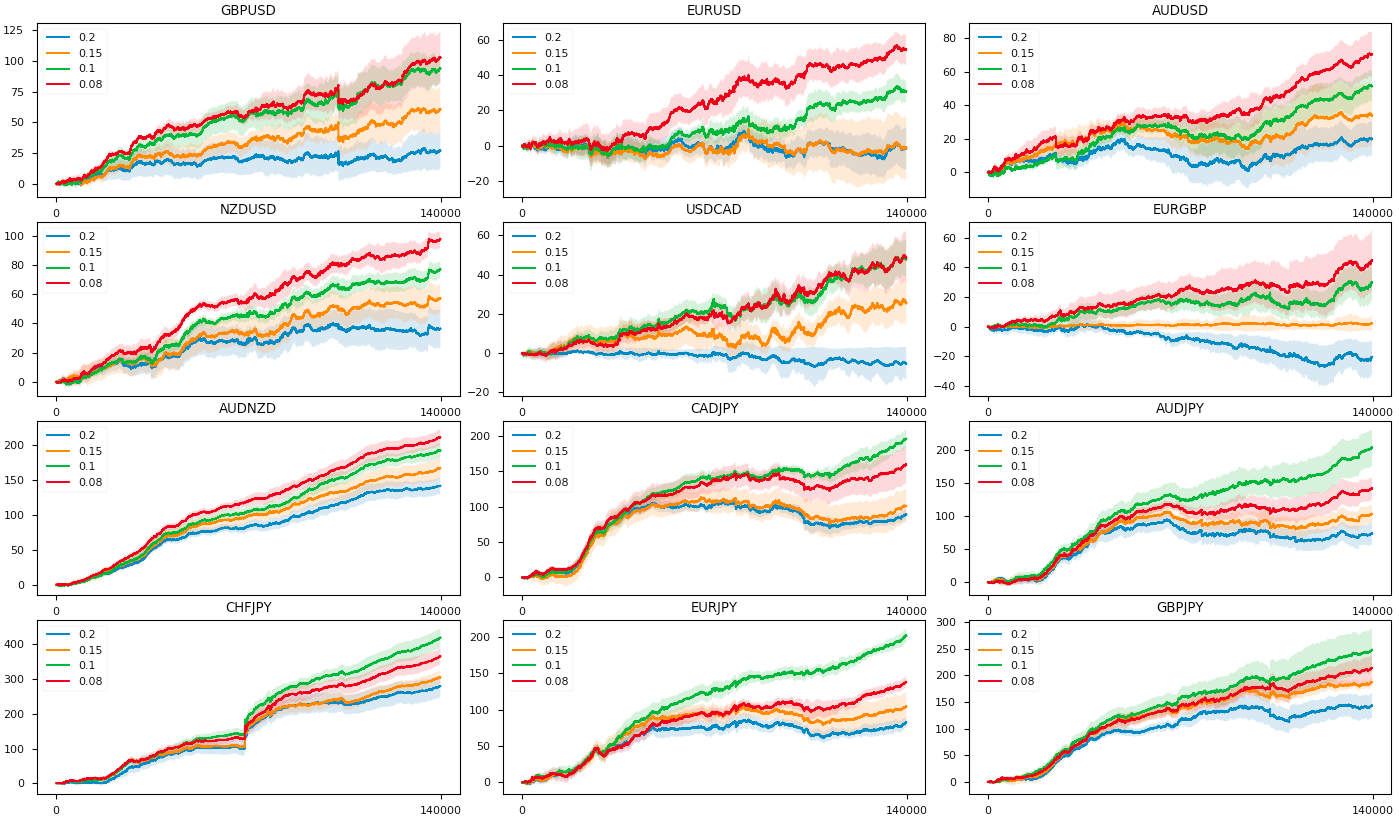}
\caption{Performances under different spreads.}
\end{figure}

\begin{figure}[h]
\centering
\includegraphics[width=.9\linewidth]{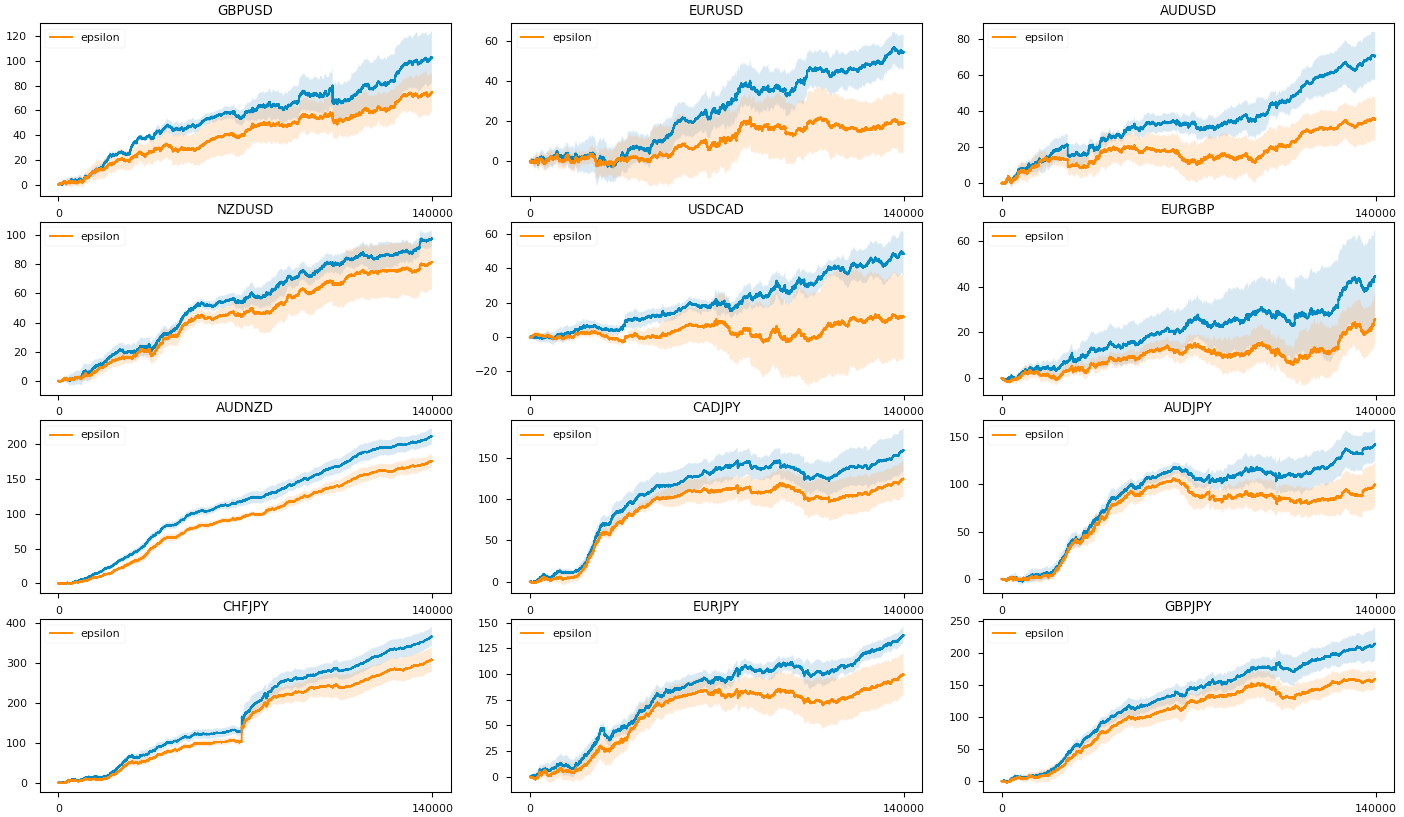}
\caption{Performances with and without action augmentation.}
\end{figure}


\newpage
\bibliography{reference}

\end{document}